\documentclass[ISTS]{tjsass} 

\usepackage[pdftex]{graphicx}
\usepackage{epstopdf}
\RequirePackage{multicol}
\RequirePackage{amsmath}
\RequirePackage[varg]{txfonts}
\RequirePackage{bm}
\RequirePackage{array}
\RequirePackage{tjsasscite}
\usepackage[hang]{footmisc}
\newcommand{\bhline}[1]{\noalign{\hrule height #1}}

\usepackage{color}
\usepackage{comment}
\usepackage[subrefformat=parens]{subcaption}
\usepackage{caption}
\pubyear{2020}
\bookvolume{17}
\bookissue{20}
\titleheadertrue

\title{Application of Machine Learning to the Particle Identification of GAPS}
\subtitle{}
\author{\NAME{Takuya}{WADA},\thanksNum{1),2)}\CorresAuthor{test@jsass.org} 
\NAME{Hideyuki}{FUKE},\thanksNum{2)}
\NAME{Yuki}{SHIMIZU},\thanksNum{3)} and
\NAME{Tetsuya}{YOSHIDA}\thanksNum{1),2)}}
\thanksOrg{Aoyama Gakuin University, Sagamihara, Japan}
\thanksOrg{Institute of Space and Astronautical Science, JAXA, Sagamihara, Japan}
\thanksOrg{Kanagawa University, Yokohama, Japan}
\begin{abstract}
GAPS is an international balloon-borne project that contributes to solving the dark-matter mystery through a highly sensitive survey of cosmic-ray antiparticles, especially undiscovered antideuterons. To achieve a sufficient sensitivity to rare antideuterons, a novel particle identification method based on exotic atom capture and decay has been developed. In parallel to utilizing this unique event signature in a conventional likelihood-based event identification scheme, we have begun investigating a complementary approach using a machine learning technique. In this new approach, a deep-learning package is trained on a large amount of input data from simulated antiparticle events through a multi-layered neural network. By applying this unbiased approach, we expect to mine unknown patterns and give feedback to the conventional method. In this paper, we report results from exploratory investigations that illustrate the promise of this new approach.
\end{abstract}
\begin{document}
\maketitle

\section{Introduction}

The origin of dark matter (DM) is a major subject for modern physics. Although the nature of DM has not yet been revealed directly, the existence of DM, which accounts for around a quarter of the total energy density of the universe, is strongly supported by many astronomical observations and theoretical calculations.\cite{bib01} A leading class of DM candidate particles is the weakly interacting massive particle (WIMP). A number of experiments have been carried out to detect DM either directly, indirectly, or using a particle accelerator. To verify a wide variety of theoretical DM models, it is important to investigate DM from diverse complementary angles.

Cosmic-ray antideuterons are expected to provide a new approach to indirectly detect DM.\cite{bib02} Antideuterons can be produced by self-annihilation or decay of WIMP DM particles, in common with the other indirect probes such as gamma rays, positrons, and antiprotons. In contrast to all these probes, the flux of DM-produced antideuterons can be orders of magnitude above the astrophysical backgrounds (originating from the secondary interactions of cosmic rays) whose abundance is kinematically suppressed in the sub-GeV low-energy region.\cite{bib03,bib04} Therefore, the detection of even a single sub-GeV antideuteron can provide evidence of a novel origin. Figure 1 shows representative antideuteron spectra predicted from DM models such as the lightest supersymmetric particle (LSP) neutralino,\cite{bib05} right-handed Kaluza-Klein neutrino of warped 5-dimensional grand unified theories (LZP),\cite{bib06} and decaying LSP gravitino.\cite{bib07} Among these models, the DM with several tens of GeV mass has been recently discussed as a possible source to interpret observed excesses of cosmic-ray antiprotons,\cite{bib09} and gamma rays.\cite{bib10} Antideuterons are still almost unexplored and have never been detected in the cosmic radiation\cite{bib13,bib14}. Hence, low-energy cosmic-ray antideuterons have a wide discovery space to detect DM.

\begin{figure}[!t]%
\centering
\includegraphics[width=85mm,clip]{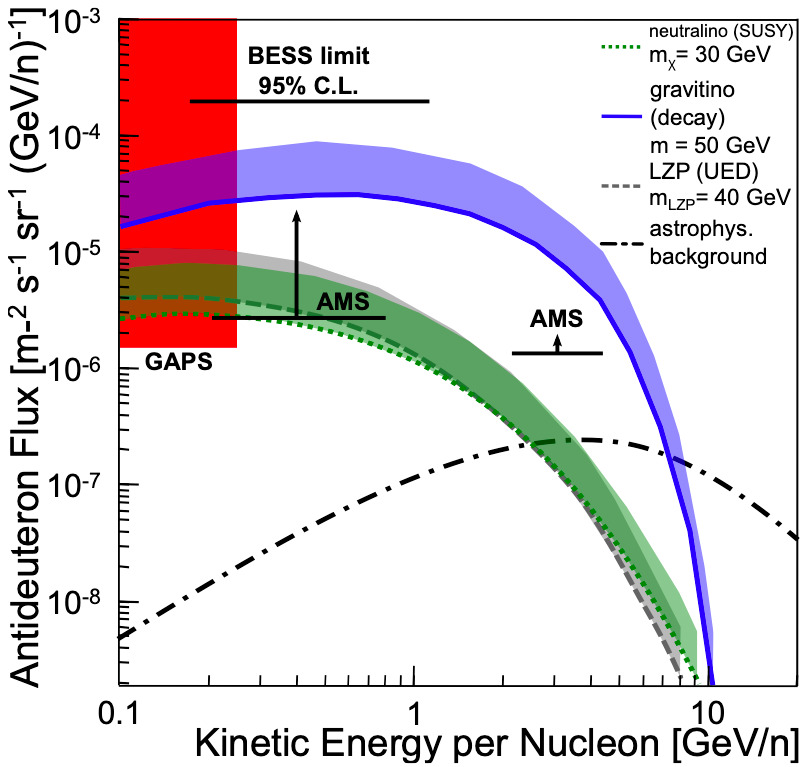}
\caption{Predicted primary antideuteron spectra at the top of the atmosphere from various dark matter models (light neutralino (green),\cite{bib05} LZP (gray),\cite{bib06} and gravitino (blue)\cite{bib07}). The bands for these spectra represent uncertainties in the absolute flux due to the typical uncertainty from the propagation calculation (MED/MAX).\cite{bib12} The astrophysical secondary/tertiary component (dashed-dotted line) is predicted to be suppressed in the low-energy region below $\sim$1 GeV/n.\cite{bib04} The GAPS antideuteron sensitivity (99\% C.L.) expected in 3 LDB flights ($\sim$105 days) is more than two orders of magnitude better\cite{bib11} than the upper limit set by the BESS experiment.\cite{bib13}.}
  \label{fig01}%
\end{figure}%

\section{GAPS Project}

The General AntiParticle Spectrometer (GAPS) is an international project to contribute to dark matter physics through a highly sensitive survey of cosmic-ray antiparticles.\cite{bib15,bib16,bib17} The primary goal of GAPS is to search for undiscovered antideuterons in the low-energy range ($<$0.25~GeV/neucleon) with an unprecedented sensitivity. To achieve a high sensitivity, GAPS plans to fly a large-grasp instrument over Antarctica multiple times by using NASA long-duration balloons (LDBs). The polar balloon flight is optimal for GAPS, not only because long observation times ($\sim$1~month) can be realized at high altitudes, but also because of the low rigidity cutoff near the geomagnetic pole, which allows us to observe charged cosmic rays directly in the low-rigidity range below $\sim$0.5~GV. These low energies are highly suppressed on the orbit of the International Space Station (ISS). The first GAPS LDB flight is planned for late 2021. The GAPS antideuteron sensitivity expected in three LDB flights is shown by Fig.~\ref{fig01}.\cite{bib11}

GAPS will also provide a precise measurement of the antiproton flux around 100~MeV, a region that is particularly sensitive to low-mass DM models.\cite{bib18} GAPS will detect more than an order of magnitude more antiprotons in the low-energy range compared to previous experiments such as BESS-Polar\cite{bib19} and PAMELA.\cite{bib20} Precise measurement in this lowest energy range offers new phase space for probing light DM models, such as light neutralinos, gravitinos, and LZPs. GAPS is also sensitive to antihelium,\cite{bib21} which is another new probe into the DM physics. However, antihelium is outside the scope of this paper.

\subsection{Detection concept}

To observe rare antiparticles among high cosmic-ray backgrounds, it is essential to survey antiparticles with a large grasp instrument also with high identification capabilities. For instance, typical fluxes of protons and antiprotons in the cosmic radiation are approximately $10^{10}$ and $10^{4}$ higher, respectively, than the antideuteron flux predicted in Fig.~\ref{fig01}.

To realize good identification capability against these backgrounds while keeping a large geometrical acceptance, GAPS introduces an original method that utilizes the deexcitation sequence of exotic atoms.\cite{bib22,bib23}

\begin{figure}[!t]%
\centering
\includegraphics[width=85mm,clip]{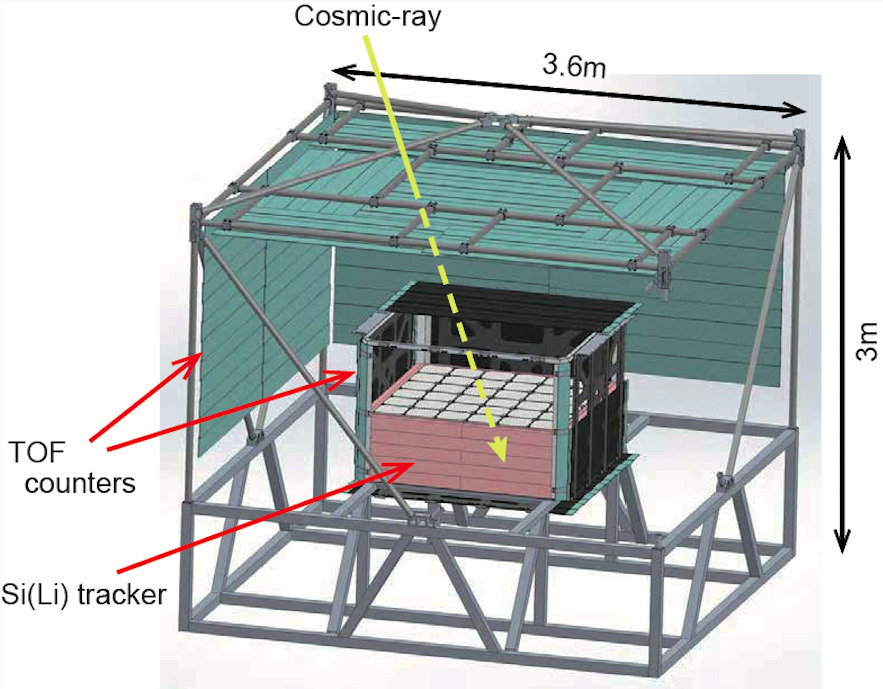}
\caption{A conceptual diagram of the GAPS instrument configuration. In the central tracker, $\sim$1000~Si (Li) detectors are arrayed in 10~Layers. The tracker is surrounded by the inner and outer TOF plastic scintillation counters. In this figure, some of Si(Li) detectors and the TOF paddles are not shown to make the Si(Li) array visible.}%
  \label{fig02}%
\end{figure}

\begin{figure}[!t]%
\centering
\includegraphics[width=85mm,clip]{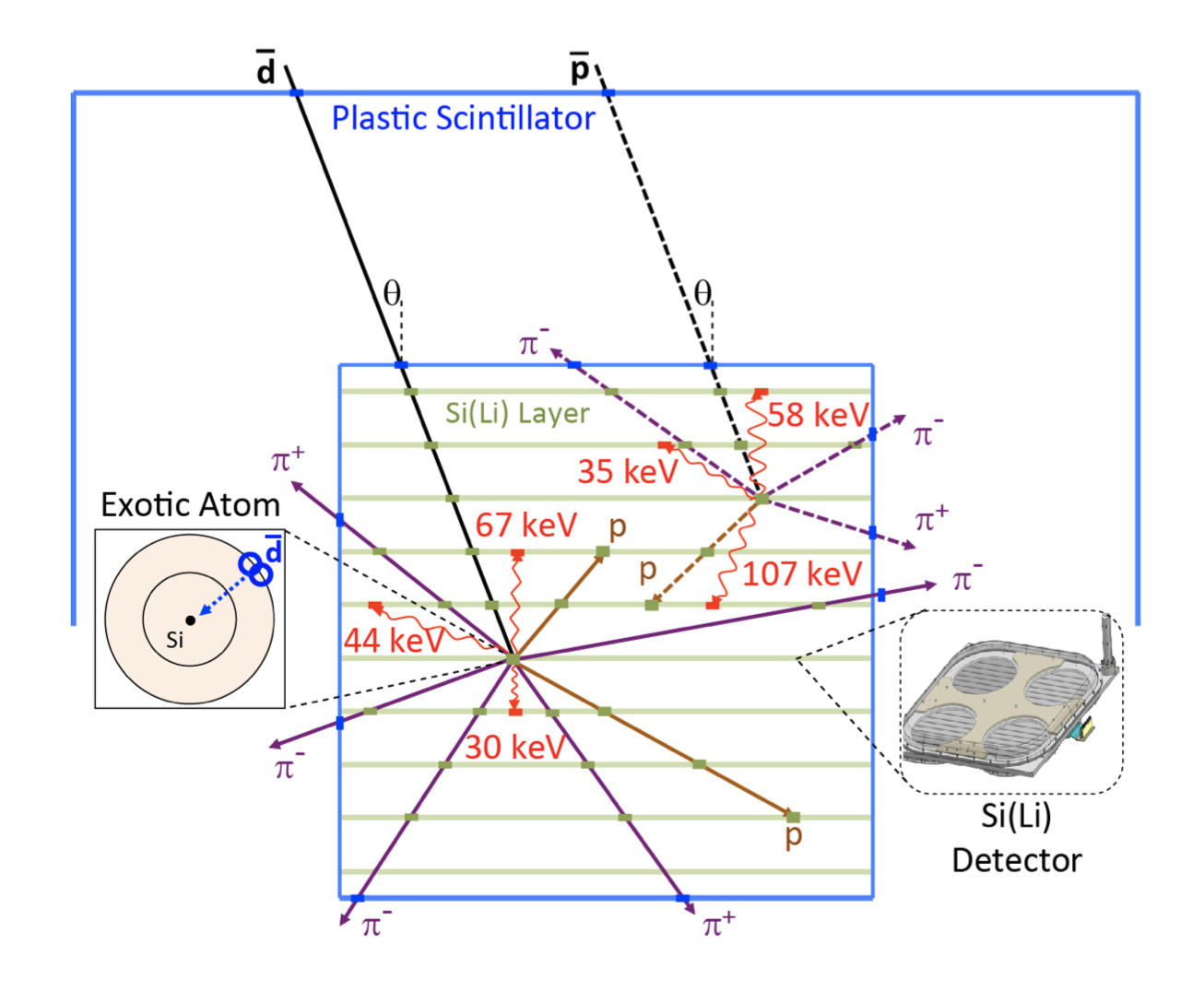}
\caption{A conceptual diagram of the GAPS antiparticle identification method. An antiparticle slows down and stops in the Si(Li) target forming an exotic atom. Through the deexcitation of the exotic atom, the characteristic X-rays will be emitted followed by the pions and protons emission in the nuclear annihilation. Using this technique, antiprotons and antideuterons in the cosmic radiation are identified with high detection efficiencies and sufficient rejection power against background events.}%
  \label{fig03}%
\end{figure}

Figure~\ref{fig02} shows a conceptual diagram of the GAPS instrument configuration. A central tracker composed of over 1000 custom lithium-drifted silicon (Si(Li)) detectors is surrounded by a time-of-flight (TOF) system. Figure~\ref{fig03} shows a conceptual diagram of the GAPS antiparticle identification method. When an antiparticle arrives from space, it is slowed down by the energy losses in the residual atmosphere, in the GAPS TOF counters, and in Si(Li) tracker as the target material. Just after stopping in the target, the antiparticle forms an exotic atom in an excited state with near unity probability. Then, through radiative transitions in the cascade to the ground state, the exotic atom deexcites with the emission of characteristic X-rays. The energies of the ladder X-rays are strictly determined by the exotic atom physics and thus provide a key to identify the incoming antiparticle species. After the X-ray emission, the antiparticle annihilates in the nucleus, emitting a characteristic number of pions and protons, which provides additional particle identification information. Tracks of X-rays and pions or protons with a vertex, in combination with other measured values such as the time-of-flight (or the velocity), the energy deposits, and the stopping depth, enables us to distinguish rare antideuterons from backgrounds including antiprotons and protons. Without the technical limitations of heavy magnets in conventional magnetic spectrometers, this technique allows us to build an instrument with a large grasp and low-energy range. The principle of this particle identification technique was verified by accelerator tests with various target materials using the KEK antiproton beam-line.\cite{bib24,bib25}

\subsection{Instrument design}

The central tracker shown in Fig.~\ref{fig02} consists of $>$1000 Si(Li) detectors arrayed in 10 layers with 10~cm vertical spacing in a 1.6~m $\times$ 1.6~m $\times$ 1~m volume. Each Si(Li) wafer has 4-inch diameter and 2.5~mm thickness and is segmented into 8 strips.\cite{bib26,bib27,bib38,bib39} The Si(Li) detector serves as a degrader, a depth sensing detector, a stopping target to form an exotic atom, an X-ray spectrometer and a charged particle tracker. In order to distinguish antideuteronic X-rays from antiprotonic X-rays, the energy resolution for X-rays should be better than $\sim$4~keV, which is achievable at operating temperatures of \mbox{$\sim$-40${}^\circ$C}.

The TOF system is composed of the inner and outer scintillation counters ($\sim$200~counters total). Each counter consists of thin ($\sim$6~mm thick) and long ($\sim$180~cm) plastic scintillator paddles whose both ends are coupled to six silicon photomultipliers (SiPMs) each. The TOF system generates the trigger signal, measures the time-of-flight, measures the energy deposit, roughly determines the arrival direction, and works as a pion/proton detector. The experimentally determined time resolution is $\sim$0.4~ns.

The basic GAPS payload design concept was successfully verified during a balloon flight in June 2012 at Taiki, Japan. Prototypes of all GAPS key components were mounted on the payload. Recording more than one million events during the flight, it was confirmed that the components including the Si(Li) detectors and TOF counters operated as expected.\cite{bib28,bib29,bib30,bib31}

\section{Approaches to Particle Identification}

Rigorous particle identification and background suppression are necessary, especially to distinguish antideuterons from backgrounds. The conventional particle identification method being developed is based on the reconstruction of each event including the incoming particle, secondary multiple particles (pions and protons), and characteristic X-rays.\cite{bib11,bib32} For the reconstruction of each event, a data set of about $10^{4}$ channels (8 strips of $\sim$a~thousand Si(Li) detectors each and both ends readout of $\sim$200~TOF counters each) are used. From each channel, energy depositions information can be provided. The TOF counters provide hit timing information, too. During the event reconstruction, key physics parameters for the particle identification are obtained. Among the backgrounds, antiprotons are considered to be the major background for antideuterons rather than the more numerous protons, because only antiparticles can form an exotic atom and fake antideuteronic signals in the GAPS identification method.\cite{bib22} From baseline studies, a sufficient background suppression capability is expected from a combination of the reconstructed physical parameters even against the antiprotons.\cite{bib11} However, it is still challenging to establish the details of the particle identification method, because of the complexity of a many-channel analysis, the large variety of expected signal patterns, and the required high identification capabilities.

Therefore, in parallel to the development of conventional particle identification methods, we have begun investigating a complementary approach using a machine learning technique. Machine learning is a subset of artificial intelligence (AI) that has found application recently in a wide variety of fields, in particular thanks to its high potential for pattern recognition.\cite{bib33} In our new approach, a deep-learning algorithm builds a mathematical model by ``learning'' on copious training data of simulated antiparticle events through a multi-layered neural network (NN) without any physical interpretations. By applying this fully unbiased approach, we aim to validate the conventional method and to mine unknown patterns. This will provide positive feedback to the conventional event identification method, further improving the GAPS antiparticle detection sensitivities. Hereafter, exploratory investigations of the deep-learning approach are discussed.

\section{Particle Identification by Machine Learning}

\subsection{Deep learning}

\begin{figure}[!t]%
\centering
\includegraphics[width=85mm,clip]{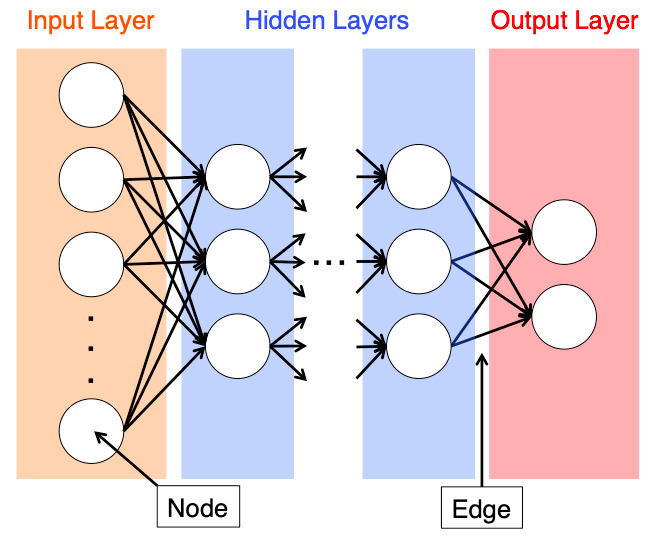}
\caption{A conceptual diagram of a deep neural network. The network is composed of an input layer, an output layer, and hidden layers. Nodes in each layer and interconnecting edges are illustrated by circles and arrows, respectively.}
  \label{fig04}%
\end{figure}

Deep learning is a machine-learning algorithm of deep NN. NN, or artificial NN, is a mathematical framework based on a collection of interconnected artificial neurons which models biological neural networks in brains. As shown by Fig.~\ref{fig04}, an NN is composed of an input layer, an output layer, and in-between hidden layers. Artificial neurons, or nodes, in each neighboring layer are interconnected by edges. Each connection is assigned a weight to adjust the connection strength. Each layer has a nonlinear activation function to compute summation outputs from the weighted inputs. The outputs produced from a layer are then inputted to nodes in the next layer. In this manner, the input values are transferred to the final layer to compute the output. The consistency between the input and the output is evaluated by a loss function. Through iterative learning processes the weights are modified so that the consistency is improved. As a result, a learning model can be obtained which produces a favored output from a given input. By using a multi-layer (or deep) NN, the recognition accuracy can be drastically improved.

\subsection{Input data}

In this study, we used 204~energy depositions from the TOF counters (one from each counters) and 11,520 energy depositions from the Si(Li) detectors (8~strips from 1,440~detectors). The input data were generated using a Monte-Carlo simulation code\cite{bib32} developed by the GAPS collaboration based on the GEANT4 framework\cite{bib34} version 10.4. In this study, we used the energy deposition values calculated by the GEANT4 code, without taking into account resolution effects. Characteristic X-rays from exotic atoms, which will be included in future GEANT4 versions, are also not taken into account in this study. Furthermore, timing measurement information is not considered.

\begin{figure}[!t]%
\centering
\includegraphics[width=85mm,clip]
{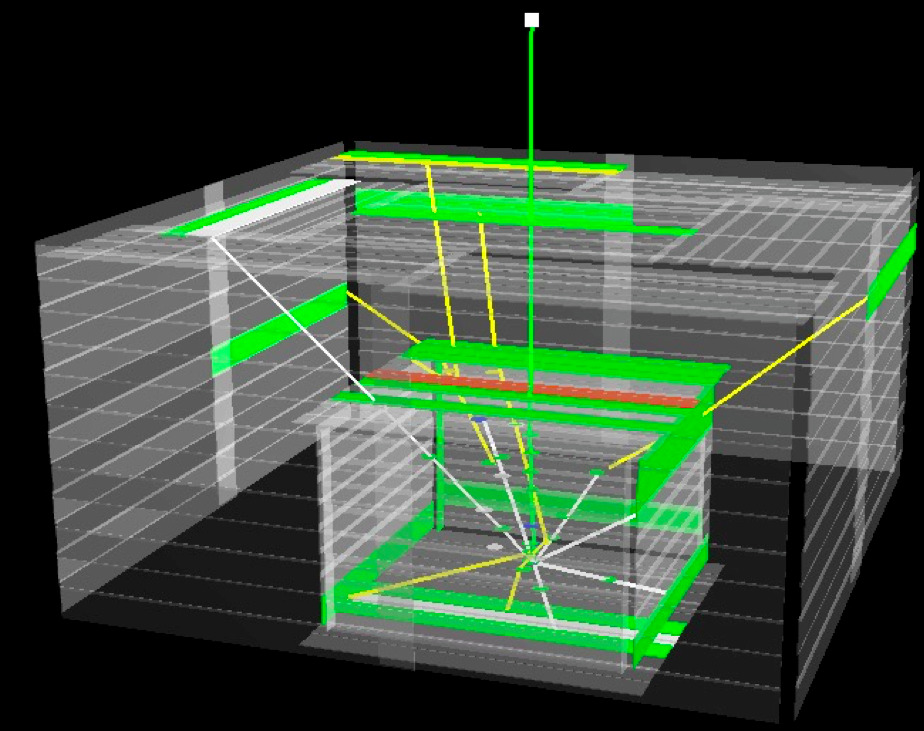}
\caption{An event display of an example of input data simulated by the GEANT4-based code. The incident particle (green line) is an antideuteron with a velocity of 0.335$<$$\beta_{1}$$<$0.340 which produces pions (white trajectories) through a nuclear annihilation in the instrument.}
  \label{fig05}%
\end{figure}

In each data set, an antideuteron or an antiproton was injected from the same position in the center above the instrument with a vertically-downward fixed incident angle (Fig.~\ref{fig05}). To simplify things, the velocity, $\beta$, of the incident antiparticle was limited to two narrow ranges; the $\beta$ was uniformly distributed by random numbers either within 0.335$<$$\beta_{1}$$<$0.340 or 0.250$<$$\beta_{2}$$<$0.255. In this study, we discuss (i) distinguishing between antideuterons and antiprotons with similar velocities of $\beta_{1}$ and (ii) distinguishing between antideuterons with different velocities of $\beta_{1}$ and $\beta_{2}$.

200,000 events data sets were prepared with the simulation code for each combination case of an antiparticle species and a velocity range. As the input of training data for the supervised learning, 160,000 labeled data sets were used for each case. The rest of 40,000 data sets were used without a label as the validation data. These number of events were limited by the amount of computing power used in this study but are be sufficient for this first-step study to explore the feasibility of our deep-learning approach.

\subsection{Learning model framework}

We used the deep-learning framework Keras,\cite{bib35} which is a Python-based open-source NN library, and a backend of TensorFlow\cite{bib36} which is supported by Keras. Table~\ref{tbl01} summarizes the outline of the network structure used. \textcolor{black}{All hidden layers were fully connected.}
\begin{table}[!t]
\centering
\caption{The outline of the network structure used in this study.}\label{tbl01}
\begin{tabular}{ccc}\bhline{0.8pt}
Layer        & Number of nodes  & Activation function  \\ \hline
Input Layer & 11724 & - \\
Hidden 1 & 8000 & ReLU \\
Hidden 2 & 4000 & ReLU \\ 
Hidden 3 & 2000 & ReLU \\
Hidden 4 & 1000 & ReLU \\
Hidden 5 & 500 & ReLU \\
Hidden 6 & 50 & ReLU \\
Output Layer  & 1 & sigmoid \\ \bhline{0.8pt}
\end{tabular}
\end{table}
As is common, a Rectified Linear Unit (ReLU or ramp function) was used as the activation function in the hidden layers. ReLU, which is expressed by Eq. (\ref{eq01}), outputs zero for negative inputs and equals input for non-negative inputs.
\begin{equation}
{\rm ReLU}\ (x) = {\rm max}(0,x) \label{eq01}.
\end{equation}
The sigmoid function used in the output layer is suitable for binary classifications like our case studies. The sigmoid function, defined by Eq.~(\ref{eq02}), computes the likelihood in a range from 0 to 1 to estimate to which class the input data should be classified.
\begin{equation}
{\rm sigmoid}\ (x) = \frac{1}{1+e^{-x}} \label{eq02}.
\end{equation}

Table~\ref{tbl02} summarizes major hyperparameters used in this study. 
\begin{table}[!t]
\centering
\caption{Hyperparameters used in this study.}\label{tbl02}
\begin{tabular}{cc}\bhline{0.8pt}
Batch size        & 320  \\ \hline
Epochs & 500 (early stopping : ON)  \\ \hline
Optimizer & Adam \\ \hline
Learning rate & 0.00001 \\ \hline
Loss function & Binary Crossentropy \\ \bhline{0.8pt}
\end{tabular}
\end{table}
Hyperparameters are the parameters that must be set before starting the learning process and are essential to control the learning algorithm. Batch size defines the number of data sets used in one iteration. Here we used the mini-batch mode so that the entire training data are divided into subsets defined by the batch size. The number of epochs defines the number of times that the learning algorithm will pass through the entire training data set. As the optimizer, which defines how to update the network weights during the iterative learning, the common gradient-based optimizer Adam\cite{bib37} was used. The learning rate defines how strongly the optimizer updates the weights, and thus controls the convergent behavior of the learning model. As the loss function we used binary cross-entropy, which is suitable to binary classifications.

The network structure and hyperparameters shown by Tables~\ref{tbl01} and \ref{tbl02} should be optimized for each learning case. Inadequate hyperparameters can result in an inaccurate model such as an overfitted model, which contains more parameters than can be justified by the data. As an option to avoid the overfitting, we incorporated the early-stopping function which terminates the learning process when the improvement of the learning accuracy saturates before reaching the number of epochs. In addition, we implement the dropout function, which is another technique to avoid the overfitting. The dropout function lowers the degree of freedom of the network by deactivating some nodes in a layer with a certain probability and improves the generalization performance. \textcolor{black}{In this study, the dropout function was applied to the input layer and all hidden layers. The dropout ratio was set to 20\% in the input layer and 50\% in each hidden layer.} Hyperparameters used in this exploratory study were tentatively chosen among various sets of values so as to achieve the highest learning accuracy. \textcolor{black}{The NN with these above hyperparamaters was commonly used for both cases of (i) and (ii) but the learning model was trained for each case independently.}

\section{Results}

\subsection{Convergence in iterative learning}

\begin{figure}[!t]%
\centering
\includegraphics[width=85mm,clip]{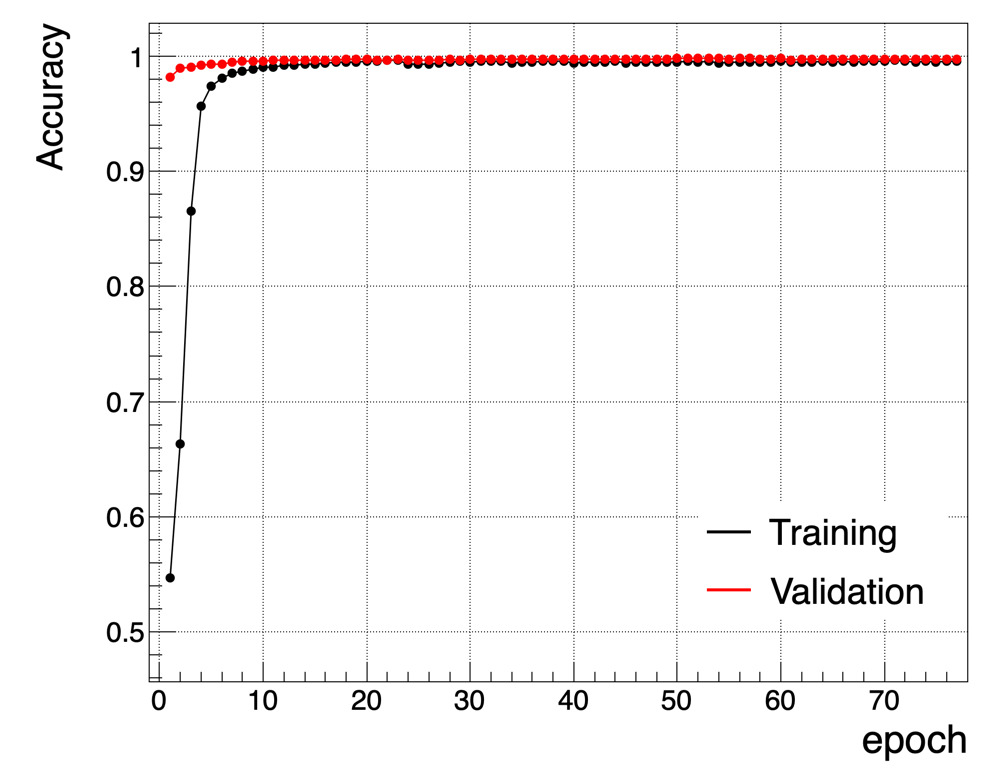}
\caption{An example of learning accuracies of training data (black) and validation data (red) as functions of the number of epochs. Both accuracies do not become worse as the number epochs becomes large, confirming that the overfitting is avoided.}
  \label{fig06}%
\end{figure}

Due to the sigmoid function (Eq.~(\ref{eq02})), the output from each input event has an output between 0 and 1. Given a binary-classification boundary threshold of $T$, each output can be judged as to which class the computed likelihood belongs. As an example, in the case that classes of ``A'' and ``B'' are tagged by the values of 1 and 0, respectively, outputs from a class-A input with a value larger than $T$ and outputs from a class-B input with a value smaller than $T$ are recognized correctly. In this manner, the recognition efficiency, $\epsilon_{recog}$, and the misidentification probability, $\epsilon_{misid}$, can be calculated as follows:
\begin{equation}
\scalebox{0.9}{$\displaystyle \epsilon_{recog,~A}\left(>T\right)$} \equiv \frac{\scalebox{0.65}{$\displaystyle number~of~data~A~recognized~as~A~with~a~value > T$}}{total~number~of~input~data~A}  \label{eq03},
\end{equation}
\begin{equation}
\scalebox{0.9}{$\displaystyle \epsilon_{misid,~A}\left(<T\right)$} \equiv \frac{\scalebox{0.65}{$\displaystyle number~of~data~A~recognized~as~B~with~a~value < T$}}{total~number~of~input~data~A}  \label{eq04},
\end{equation}
\begin{equation}
\scalebox{0.9}{$\displaystyle \epsilon_{recog,~B}\left(<T\right)$} \equiv \frac{\scalebox{0.65}{$\displaystyle number~of~data~B~recognized~as~B~with~a~value < T$}}{total~number~of~input~data~B}  \label{eq05},
\end{equation}
\begin{equation}
\scalebox{0.9}{$\displaystyle \epsilon_{misid,~B}\left(>T\right)$} \equiv \frac{\scalebox{0.65}{$\displaystyle number~of~data~B~recognized~as~A~with~a~value > T$}}{total~number~of~input~data~B}  \label{eq06}.
\end{equation}
A commonly used index of accuracy is defined by the total rate of correct outputs with $T$ = 0.5:
\begin{equation}
\scalebox{0.8}{$\displaystyle Accuracy$} \equiv \frac{
\scalebox{0.8}{$\displaystyle \parbox{22em}{($number~of~data~A~with~a~value > 0.5 \\ ~~~~~~~~~~~~~~~~~~~plus~number~of~data~B~with~a~value < 0.5$)}$}
}{total~number~of~input~data~A~and~B}  \label{eq07}.
\end{equation}
As an example of the learning curve, Fig.~\ref{fig06} shows the accuracy profile in the case of distinguishing antiprotons and antideuterons with similar velocities of $\beta_{1}$. \textcolor{black}{The high accuracies at the first few epochs validate that the hyperparameters were set appropriately.} Accuracies both of the training and validation data are improved by the iterative learning and converge to 1. This confirms that the overfitting is successfully avoided. In every case, we confirmed the convergence in this manner. The accuracy of the training data in this plot is lower than that of the validation data because the training-data accuracy is underestimated due to the dropout function.

For discussions in the following section, here we also define the rejection power, $p_{reject}$, as follows;
\begin{equation}
p_{reject,~A}\left(>T\right) \equiv \frac{1}{\epsilon_{misid}\left(A,<T\right)} \label{eq10},
\end{equation}
\begin{equation}
p_{reject,~B}\left(<T\right) \equiv \frac{1}{\epsilon_{misid}\left(B,<T\right)} \label{eq11}.
\end{equation}


\subsection{Distinguishing between antideuteron and antiproton with a similar velocity}

\begin{figure}[!t]%
\centering
\includegraphics[width=85mm,clip]{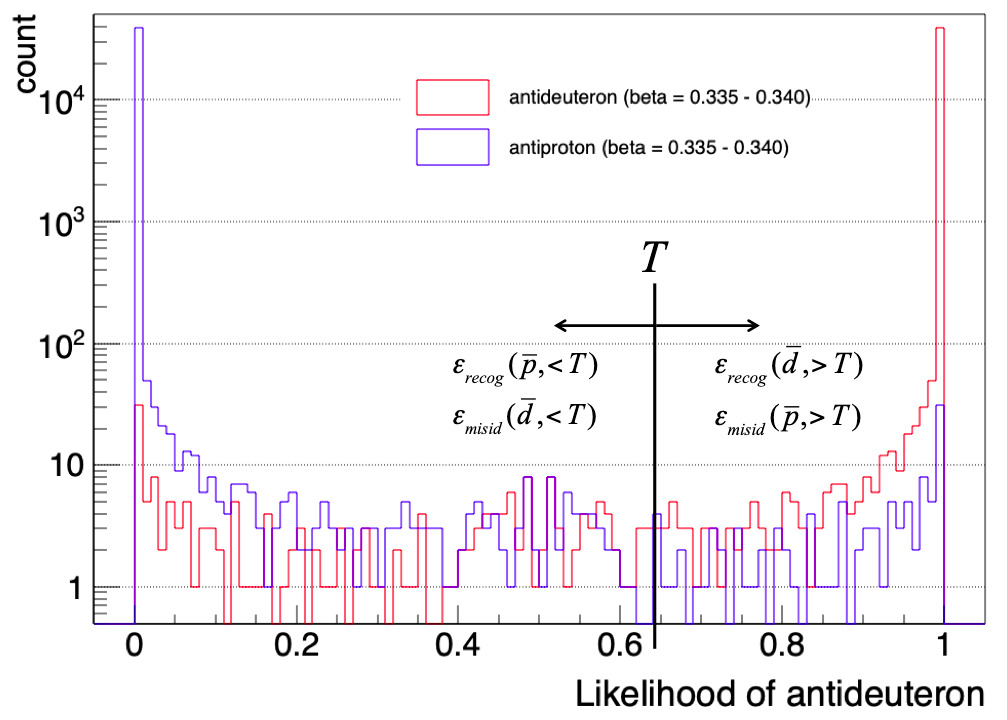}
\caption{The output likelihood distributions in semi-log format. Most of values calculated from the antideuteron validation data (red) and antiproton validation data (blue) are recognized correctly.}
  \label{fig07}%
\end{figure}

\begin{figure*}[!t]
 \begin{minipage}[b]{0.5\linewidth}
  \centering
  \includegraphics[keepaspectratio, scale=0.36]
  {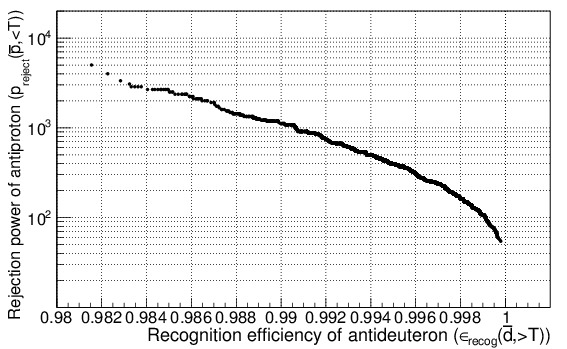}
  \subcaption{}\label{fig09a}
 \end{minipage}
 \begin{minipage}[b]{0.5\linewidth}
  \centering
  \includegraphics[keepaspectratio, scale=0.36]
  {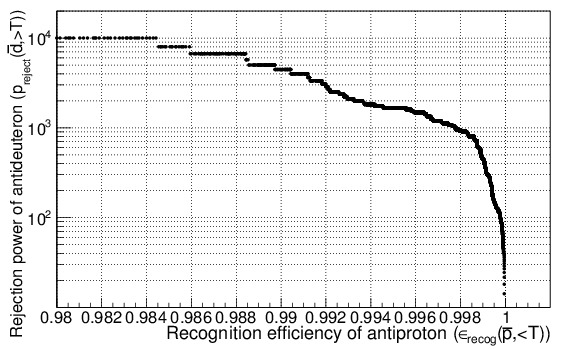}
  \subcaption{}\label{fig09b}
 \end{minipage}
 \caption{\subref{fig09a} Rejection power of $\beta_{1}$ antiprotons against recognition efficiency of $\beta_{1}$ antideuterons. \subref{fig09b} Rejection power of $\beta_{1}$ antideuterons against recognition efficiency of $\beta_{1}$ antiprotons.}\label{fig09}
\end{figure*}

Figure~\ref{fig07} shows the output likelihood distributions calculated from the validation data in the case of distinguishing antideuterons and antiprotons with velocities of $\beta_{1}$. Most of antideuterons (red) and antiprotons (blue) are correctly recognized. By varying the threshold $T$, the $\epsilon_{recog}$ of true input can be plotted.
Figure~\ref{fig09}\subref{fig09a} shows the relation between the $\epsilon_{recog}$ of $\beta_{1}$ antideuteron and $p_{reject}$ of $\beta_{1}$ antiproton. The $p_{reject}$ reaches above $10^{3}$ while keeping a high $\epsilon_{recog}$ of $\sim$\nolinebreak98\%.

In the inverse case of recognizing antiproton, as shown by Fig.~\ref{fig09}\subref{fig09b}, rejection power well above $10^{3}$ is achieved while keeping a high identification efficiency of $\sim$99\%.

\subsection{Distinguishing between antideuterons in two velocity ranges}

\begin{figure*}[!t]
 \begin{minipage}[b]{0.5\linewidth}
  \centering
  \includegraphics[keepaspectratio, scale=0.36]
  {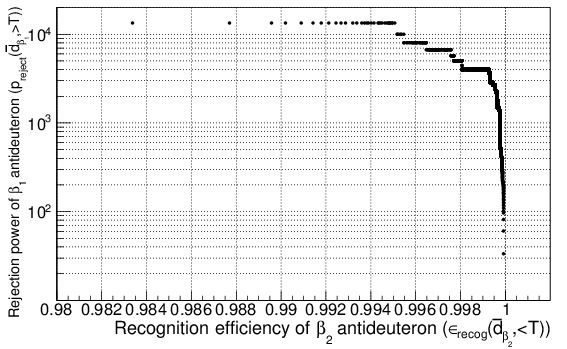}
  \subcaption{}\label{fig11a}
 \end{minipage}
 \begin{minipage}[b]{0.5\linewidth}
  \centering
  \includegraphics[keepaspectratio, scale=0.36]
  {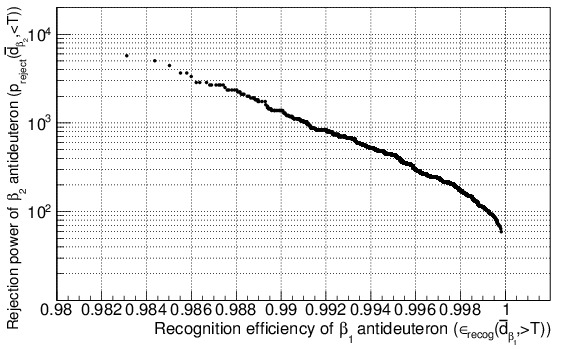}
  \subcaption{}\label{fig11b}
 \end{minipage}
 \caption{\subref{fig11a} Rejection power of $\beta_{1}$ antideuterons against recognition efficiency of $\beta_{2}$ antideuterons. \subref{fig11b} Rejection power of $\beta_{2}$ antideuterons against recognition efficiency of $\beta_{1}$ antideuterons.}\label{fig11}
\end{figure*}

Figure~\ref{fig11} shows the $p_{reject}$ as functions of the $\epsilon_{recog}$ in the case of distinguishing antideuterons with velocities of 0.335$<$$\beta_{1}$$<$0.340 and 0.250$<$$\beta_{2}$$<$0.255. High rejection powers above $10^{3}$ are achieved while keeping a high identification efficiency of $\sim$\nolinebreak99\%.

\subsection{Discussions}

From the results, in each case, high rejection powers of $\sim$\nolinebreak$10^{3}$ are achieved while keeping high identification efficiencies above $\sim$\nolinebreak98\%. This indicates the potential of the deep-learning approach to study the particle identification capability of the GAPS instrument. Indeed, in this study, the highest rejection power of $\sim$\nolinebreak$10^{4}$ achieved by Figs.~\ref{fig09} and \ref{fig11} are limited by the number of validation data of $\sim$\nolinebreak$10^{4}$. By increasing the number of both the training and validation data, the learning accuracy will be increased.

In this study, characteristic X-rays\cite{bib11} and timing measurements are not included in the simulated data. By implementing these information, the distinguishing accuracy must be further increased.


\section{Conclusion}

We have begun a study using up-to-date machine learning techniques for the GAPS particle identification. These exploratory investigations indicate that this new approach can achieve a high recognition efficiency while keeping a sufficient rejection power even without using timing and characteristic information. Based on these encouraging first-step results, we will proceed full-scale studies under more realistic conditions; for instance, we will randomize incident positions, angles, and velocities of the incoming particles, input values with finite measurement resolutions, and increase statistics of both training and validation data. Thereby expanding and optimizing of the deep learning neural network will be pursued. Also, mining of unknown patterns and their feedback to the conventional identification method will be studied.


\section*{Acknowledgments}\label{Acknowledgments}

This work is partly supported by Grants-in-aid KAKENHI (JP17H01136, JP19H05198) and the Sumitomo Foundation fiscal 2018 grant for basic science research projects.


\begin{thebibliography}{99}
\bibitem{bib01}
  Klasen, M., Pohl, M., and Sigl, G.: Indirect and Direct Search for Dark Matter, \textit{Prog. in Particle and Nucl. Phys.}, \textbf{85} (2015), pp.1--32.
\bibitem{bib02}
  Aramaki, T., Boggs, S., Bufalino, S., Dal, L., Doetinchem, v. P., Donate, F., et al.: Review of the Theoretical and Experimental Status of Dark Matter Identification with Cosmic-ray Antideuterons, \textit{Phys. Rep.}, \textbf{618} (2016) pp. 1--37.
\bibitem{bib03}
  Donato, F., Fornengo, N., and Salati, P.: Antideuterons as a Signature of Supersymmetric Dark Matter, \textit{Phys. Rev. D}, \textbf{62} (2000), 043003.
\bibitem{bib04}
  Ibarra, A. and Wild, S.: Determination of the Cosmic Antideuteron Flux in a Monte Carlo Approach, \textit{Phys. Rev. D}, \textbf{88} (2013), 023014.
\bibitem{bib05}
  Donato, F., Fornengo, N., and Maurin, D.: Antideuteron Fluxes from Dark Matter Annihilation in Diffusion models, \textit{Phys. Rev. D}, \textbf{78} (2008), 043506.
\bibitem{bib06}
  Baer H. and Profumo, S.: Low Energy Antideuterons: Shedding Light on Dark Matter, \textit{J. of Cosmology and Astroparticle Phys.}, \textbf{0512} (2005), 008.
\bibitem{bib07}
Dal, A. L. and Raklev, R. A.: Antideuteron Limits on Decaying Dark Matter with a Tuned Formation Model, \textit{Phys. Rev. D}, \textbf{89} (2014), 103504.
\bibitem{bib09}
  Cui, M.Y., Yuan, Q. Tsai, S. Y.K., and Fan, Y.Z.: Possible Dark Matter Annihilation Signal in the AMS-02 Antiproton Data, \textit{Phys. Rev. Lett.}, \textbf{118} (2017) 191101.
\bibitem{bib10}
  Daylan, T., Finkbeiner, P. D., Hooper, D., Linden, T., Portillo, K.N. S., Rodd, L. N., et al.: The Characterization of the Gamma-ray Signal from the Central Milky Way: A Case for Annihilating Dark Matter, \textit{Phys. of the Dark Universe}, \textbf{12} (2016), pp. 1--23.
\bibitem{bib11}
  Aramaki, T., Hailey, J. C., Boggs, E. S., Doetinchem, v. P., Fuke, H., Mognet, I. S., et al.: Antideuteron Sensitivity for the GAPS Experiment, \textit{Astroparticle Phys.}, \textbf{74} (2016), pp. 6--13.
\bibitem{bib12}
  Giesen, G., Boudaud, M., Genolini, Y., Poulin, V., Cirelli, M., Salati, P., et al.: AMS-02 Antiprotons, at Last! Secondary Astrophysical Component and Immediate Implications for Dark Matter, \textit{J. of Cosmology and Astroparticle Phys.}, \textbf{1509} (2015), 023.
\bibitem{bib13}
  Fuke, H., Maeno, T., Abe, K., Haino, S., Makida, Y., Matsuda S., et al.: Search for Cosmic-ray Antideuterons, \textit{Phys. Rev, Lett.}, \textbf{95} (2005), 081101.
\bibitem{bib14}
  Sakai, K., BESS collaboration: Search for Cosmic-ray Antideuterons with BESS-Polar II, 2nd Cosmic-ray Antideuteron Workshop, Los Angeles, Mar. 27th 2019 (oral presentation).
\bibitem{bib15}
  Hailey, J. C., GAPS collaboration: An Indirect Search for Dark Matter using Antideuterons: the GAPS Experiment, \textit{New J. of Phys.}, \textbf{11} (2009), 105022.
\bibitem{bib16}
  Ong, A. R., Aramaki, T., Bird, R., Boezio, M., Boggs, E. S., Carr, R., et al.: The GAPS Experiment to Search for Dark Matter using Low-energy Antimatter, Proceedings of 35th Intl. Cosmic Ray Conference (ICRC2017), Busan, Korea, 914, 2017.
\bibitem{bib17}
  Fuke, H., Aramaki, T., Boggs, S., Craig, W. W., Doetinchem, v. P., Fabris, L., et al.: Present Status and Future Plans of GAPS Antiproton and Antideuteron Measurement for Indirect Dark Matter Search, \textit{Phys. Society of Japan Conf. Proc.}, \textbf{18} (2017), 011003.
\bibitem{bib18}
  Aramaki, T., Boggs, S.E., Doetinchem. P.v., Fuke, H., Hailey, C.J., Mognet, I. S., et al.: Potential for Precision Measurement of Low-energy Antiprotons with GAPS for Dark Matter and Primordial Black Hole Physics, \textit{Astroparticle Phys.}, \textbf{59} (2014), pp. 12--17.
\bibitem{bib19}
  Abe, K., Fuke, H., Haino, S., Hams, T., Hasegawa, M., Horikoshi, A., et al.: Measurement of the Cosmic-Ray Antiproton Spectrum at Solar Minimum with a Long-Duration Balloon Flight over Antarctica, \textit{Phys. Rev. Lett.}, \textbf{108} (2012), 051102.
\bibitem{bib20}
  Adriani, O., Barbarino, C. G., Bazilevskaya, A. G., Bellotti, R., Boezio, M., Bogomolov, A. E., et al.: PAMELA Results on the Cosmic-Ray Antiproton Flux from 60 MeV to 180 GeV in Kinetic Energy, \textit{Phys. Rev. Lett.}, \textbf{105} (2010), 121101.
\bibitem{bib21}
  Carlson, E., Coogan, A., Linden, T., Profumo, S., Ibarra, A., and Wild, S. : Antihelium from Dark Matter, \textit{Phys. Rev. D}, \textbf{89} (2014), 076005.
\bibitem{bib22}
  Mori, K., Hailey, J. C., Baltz, A. E., Craig, W. W., Kamionkowski, M., Serber, T. W., et al.: A Novel Antimatter Detector based on X-ray Deexcitation of Exotic Atoms, \textit{Astrophys. J.}, \textbf{566} (2002), pp. 604--616.
\bibitem{bib23}
  Hailey, J. C., Craig, W. W., Harrison, A. F., Hong, J., Mori, K., Koglin, J., et al.: Development of the Gaseous Antiparticle Spectrometer for Space-based Antimatter Detection, \textit{Nucl. Instr. and Methods B}, \textbf{214} (2004), pp. 122--125.
\bibitem{bib24}
  Hailey, J. C., Aramaki, T., Craig, W. W., Fabris, L., Gahbauer, F., Koglin, J., et al.: Accelerator Testing of the General Antiparticle Spectrometer; a Novel Approach to Indirect Dark Matter Detection, \textit{J. of Cosmology and Astroparticle Phys.}, \textbf{0601} (2006), 007.
\bibitem{bib25}
  Aramaki, T., Chan, K. S., Craig, W. W., Fabris, L., Gahbauer, F., Hailey, J. C., et al.: A Measurement of Atomic X-ray Yields in Exotic Atoms and Implications for an Antideuteron-based Dark Matter Search, \textit{Astroparticle Phys.}, \textbf{49} (2013), pp. 52--62.
\bibitem{bib26}
  Perez, K., Aramaki, T., Hailey, J. C., Carr, R., Erjavec, T., Fuke, H., et al.: Fabrication of Low-cost, Large-area Prototype Si(Li) Detectors for the GAPS Experiment, \textit{Nucl. Instr. and Methods A}, \textbf{905} (2018), pp. 12--21.
\bibitem{bib38}
  Kozai, M., Fuke, H., Yamada, M., Perez, K., Erjavec, T., Hailey, J. C., et al.: Developing a Mass-production Model of Large-area Si(Li) Detectors with High Operating Temperature, arXiv:1906.05577
\bibitem{bib39}
  Rogers, F., Xiao, M., Perez, M. K., Boggs, S., Erjavec, T., Fabris, L., et al.: Large-area Si(Li) Detectors for X-ray Spectrometry and Particle Tracking in the GAPS Experiment, arXiv:1906.00054
\bibitem{bib27}
  Kozai, M., Fuke, H., Yamada, M., Erjavec, T., Hailey, J. C., Kato, C., et al.: Development of Large-area Lithium-drifted Silicon Detectors for the GAPS Experiment, Proceedings of 2018 IEEE Nucl. Science Symposium and Medical Imaging Conference (NSS/MIC), Sydney, Australia, N-17-01, 2018.
\bibitem{bib28}
  Fuke, H., Ong, A. R., Aramaki, T., Bando, N., Boggs, E. S., Doetinchem, v. P., et al.: The pGAPS Experiment: an Engineering Balloon Flight of Prototype GAPS, \textit{Adv. Space Res.}, \textbf{53} (2014), pp. 1432--1437.
\bibitem{bib29}
  Mognet, I. S., Aramaki, T., Bando, N., Boggs, E. S., Doetinchem, v. P., Fuke, H., et al.: The Prototype (pGAPS) Experiment, \textit{Nucl. Instr. and Methods A}, \textbf{735} (2014), pp. 24--38.
\bibitem{bib30}
  Doetnchem, v. P., Aramaki, T., Bando, N., Boggs, E. S., Fuke, H., Gahbauer, H. F., et al.: The Flight of the GAPS Prototype Experiment, \textit{Astroparticle Phys.}, \textbf{54} (2014), pp. 93--109.
\bibitem{bib31}
  Fuke, H., Okazaki, S., Ogawa, H., and Miyazaki, Y.: Balloon Flight Demonstration of an Oscillating Heat Pipe, \textit{J. or Astronomical Instr.}, \textbf{6}(2) (2017), 1740006.
\bibitem{bib32}
  Stoessl, A., GAPS collaboration: Status of the GAPS Simulation and Analysis Development, 2nd Cosmic-ray Antideuteron Workshop, Los Angeles, Mar. 27th 2019 (oral presentation).
\bibitem{bib33}
  LeCun, Y., Bengio, Y., and Hinton, G.: Deep Learning, \textit{Nature}, \textbf{521} (2015), pp. 436--444.
\bibitem{bib34}
  Agostinelli, S., Allison, J., Amako, K., Apostolakis, J., Araujo, H., Arce, P., et al.: GEANT4 - a Simulation Toolkit, \textit{Nucl. Instr. and Methods A}, \textbf{506} (2003), pp. 250--303.
\bibitem{bib35}
  Chollet, F., and others.: Keras (2015), https://keras.io. (accessed Apr. 10th 2019)
\bibitem{bib36}
  Abadi, M., Barham, P., Chen, J., Chen, Z., Davis, A., Dean, J., et al.: TensorFlow: A System for Large-Scale Machine Learning, Proceedings of 12th USENIX Symp.  Operating System Design and Implementation (OSDI’16), Savannah, GA, USA, pp. 265--283, 2016; arXiv:1605.08695.
\bibitem{bib37}
  Kingma, D.P. and Ba, J.L., Adam: a Method for Stochastic Optimization, Proceedings of 3rd Intl. Conf. for Learning Representations, San Diego, USA, 2015; arXiv:1412.6980.
\end{thebibliography}
\end{document}